\documentclass{aa}

\usepackage{graphicx}
\usepackage{xcolor}
\usepackage{booktabs}
\usepackage{amsmath}
\usepackage{txfonts}
\usepackage{lipsum}
\usepackage{subcaption}        
\usepackage{lscape}           
\usepackage{placeins} 
\usepackage[normalem]{ulem}
\usepackage[colorlinks=true, citecolor=blue, linkcolor=blue, urlcolor=blue]{hyperref}

\begin{document}

\title{Surveying for close binaries among lithium-rich giants \thanks{Based on observations collected at the European Southern Observatory under ESO programmes P116.2929.001 and P116.2929.002.}}

\author{Jesús Torres-Sánchez\inst{1}\corrauth{jetorres2@uc.cl}
     \and Sebastián Vilaza-Dallago\inst{2}\email{s.j.vilazadallago@uva.nl} \and Julio Chanamé\inst{1}\email{jchaname@astro.puc.cl} \and Rafael Brahm\inst{3}\email{rbrahm@gmail.com}
     }

\institute{Instituto de Astrofísica, Pontificia Universidad Católica de Chile, Av. Vicuña Mackenna 4860, 782-0436, Macul, Santiago, Chile \and Anton Pannekoek Institute for Astronomy, University of Amsterdam, Science Park 904, 1098 XH Amsterdam, The Netherlands \and Facultad de Ingeniería y Ciencias, Universidad Adolfo Ibáñez, Av. Diagonal las Torres 2640, 7941169 Peñalolén, Santiago, Chile}

\date{}
   \abstract 
   {The existence of low-mass giants with high lithium (Li) abundances poses a challenge to our understanding of post-main sequence stellar structure and evolution. Li-rich giants are preferentially found in the red clump (RC) rather than on the first-ascent red giant branch (RGB). One of the scenarios proposed to explain this Li enrichment involves tidal interaction with a close binary companion, which, if dominant, predicts an enhanced binary fraction among Li-rich giants relative to Li-normal ones.}
   {We aim to test the binary interaction scenario for the origin of Li-rich giants by using multi-epoch radial velocities (RVs) in order to measure the close binary fraction of a statistically robust sample of Li-rich giants.  The surveyed sample contains both RC and RGB targets, which allows us to examine potential differences in the binary fraction as a function of evolutionary stage.}
   {We monitored 70 giants (40 RC and 30 RGB) using the ESPRESSO spectrograph, 57 (32 RC and 25 RGB) of which already had four or more visits over a $\sim$7-month baseline.  The analysis presented here is based on these latter targets.  We first used a criterion based on surface gravity and the maximum observed RV difference within our data in order to discard RV variations attributable to intrinsic stellar jitter.  For the targets with RV variation larger than this threshold, we performed Monte Carlo simulations over a grid of companion mass and orbital period, to estimate the probability that the observed RV variation is due to a stellar ($P_{SB}$) or substellar ($P_{SS}$) companion.}
   {After removing targets consistent with stellar jitter, 11 ($34 \pm 8$\%) RC and eight ($32 \pm 9$\%) RGB stars remained as binary candidates or stars with a companion. Of these, three targets ($16 \pm 8$\% of the filtered sample) are confirmed stellar binaries. The distributions of $P_{SB}$ and $P_{SS}$ are statistically indistinguishable between the RC and RGB samples.}
   {We find no evidence in our current data for a binary fraction for Li-rich giants (in either evolutionary stage) larger than that expected for a regular Li-normal population, which suggests that binary interaction is unlikely to be the dominant mechanism driving Li enrichment in this sample. Extending the RV baseline to probe longer orbital periods will be important to continue exploring the plausibility of this scenario.}


\maketitle
\nolinenumbers

\section{Introduction} \label{sec:intro}
According to stellar evolution theory, as a star leaves the main sequence and ascends the red giant branch (RGB), the convective envelope deepens, dredging up (i.e., the first dredge-up, FDU) material from the inner layers---where the elemental composition has been altered by nuclear reactions---to the surface, thereby diluting the abundances of several elements \citep{1965ApJ...142.1447I}. Lithium (Li) constitutes one of the clearest examples of this phenomenon. As a fragile element that cannot be easily replenished, its abundance in stellar atmospheres undergoes dramatic changes over the course of a star's lifetime, making it a sensitive indicator of stellar evolution. In stars with masses similar to the Sun, Li is expected to be destroyed solely via proton-capture reactions; however, during the FDU, material from deep internal layers is mixed into regions hot enough to further destroy Li, compounding the dilution already induced by the mixing itself \citep{1967ARA&A...5..571I}.

Therefore, given the discovery of the first Li-rich giant star by \citet{1982ApJ...255..577W}, a new mystery began, for which a complete explanation has not appeared until now. Since then, there exist a growing number of giant stars with Li abundances that are near to, or exceed, Big Bang nucleosynthesis predictions. The discoveries of additional members of this population have increased remarkably due to large spectroscopic surveys, such as the Large Sky Area Multi-Object Fiber Spectroscopic Telescope (LAMOST) \citep{2012RAA....12.1197C, 2019ApJS..245...33G}, GALactic Archaeology with High Efficiency and Resolution Multi-Element Spectrograph (GALAH, HERMES) \citep{2019MNRAS.484.2000D, 2021MNRAS.505.5340M} and \textit{Gaia}-European Southern Observatory (ESO) survey \citep{2021A&A...651A..84M}, among others \citep{2016MNRAS.461.3336C, 2018A&A...617A...4S, 2019ApJ...880..125C, 2024ApJ...964...42S}. Standard models of stellar evolution predict an upper limit of Li abundance of $A(\mathrm{Li})= 1.5$\,dex\footnote{$A(\mathrm{Li})=\log[N(\mathrm{Li})/N(\mathrm{H})]+12$}, meaning that all the giant stars with $A(\mathrm{Li}) \geq 1.5$ are defined as Li-enriched giants. We note that this threshold is largely conventional rather than derived from a single, universal theoretical limit; the actual expected Li abundance depends on several factors, most notably stellar mass, with more massive giants able to sustain higher ``normal'' Li abundances \citep{2016ApJ...829..127A, chaname2022mass, 2023AJ....166...60T}.  Nevertheless, here we adopt the traditional limit of $A(\mathrm{Li})= 1.5$\,dex which is the typical post dredge-up value predicted by standard evolution for Population I stars. There are even stars named super Li-rich giants \citep{2014A&A...564L...6M, 2021ApJ...913L...4S}, in which the Li abundances can even exceed the meteoric $A(\mathrm{Li})= 3.3$\,dex \citep{lodders1998planetary}, which is representative of the abundance of the Solar System at its formation. Although by now several thousands of Li-rich giant stars have been detected, they still represent only $1$-$2\%$ of the giants \citep{2019MNRAS.484.2000D} and are even rarer in globular clusters \citep{2020A&A...639A...2P}, but the absence of an explanation for the formation of these stars represents a challenge to standard stellar evolution models.

In response to the observational data, several mechanisms have been proposed to explain the acquisition or production of Li in evolved stars. On the one hand, internal Li production is possible through the Cameron-Fowler mechanism \citep{1971ApJ...164..111C}, where helium (He) isotopes, $^{3}$He and $^{4}$He, must fuse together at high temperatures to produce beryllium-7 ($^{7}$Be) via proton-proton reaction and then $^{7}$Be has to be rapidly transported to cooler regions of the stellar surface in order to decay via electron capture where it can survive without being immediately destroyed by proton capture; these requirements make Li extremely sensitive to the structure and mixing inside a star. In this context, several extra-mixing mechanisms have been considered as ways to transport freshly produced $^{7}$Be to the surface, such as thermohaline mixing \citep{2008ApJ...677..581E, 2010A&A...522A..10C}, rotation-induced mixing \citep{2004ApJ...612.1081D}, and magnetic buoyancy \citep{2008ApJ...684L..29N}. We note, however, that the thermohaline models of \citet{2010A&A...522A..10C} predict net Li destruction of the RGB, and that 3D simulations indicate that the purely hydrodynamic thermohaline mixing is weaker than assumed in early prescriptions \citep{Brown+13, Fraser+26}. Magnetic fields or rotation may enhance it \citep{Sengupta+18,Harrington+19, Fraser+24}, but it has not been demonstrated whether this can lead to Li production. On the other hand, an external source might increase the surface Li abundance, such as the ingestion of planets or substellar objects by the red giant's external convective zone \citep{1999MNRAS.308.1133S, 2012ApJ...757..109C, 2016ApJ...829..127A, 2022A&A...657A..33A, 2023A&A...670A..73A, 2020ApJ...889...45S} or mass transfer from an asymptotic giant branch (AGB) star that can produce Li by hot bottom burning \citep{1992ApJ...392L..71S}. It is even possible that a combination of an internal and external Li production mechanism, for example, a mixing induced by tidal spin-up driven by a binary companion that warms the burning regions, could produce the necessary levels of increased mixing to transport internally-produced Li to the surface \citep{2019ApJ...880..125C}.

Given that the different mechanisms do not necessarily operate the same way throughout the evolutionary stage at which Li-rich giants are found, it is particularly important to establish if there is a correlation between the evolutionary stage and the Li enrichment. Different studies \citep[e.g.,][]{2019ApJ...878L..21S, 2021NatAs...5...86Y} have found that most low-mass ($M\leq 2M_\odot$) Li-rich giants lie in the core helium-burning phase (i.e., the red clump, RC). This has led to the suggestion that the enrichment mechanism could be tied to the helium flash or to the upper RGB. However, this trend is not absolute; not every Li-rich giant belongs to the RC and the distribution of Li abundances differs between both evolutionary phases \citep{2021MNRAS.505.5340M}, hinting that different mechanisms might be at play depending on the stage of the star. This distinction between evolutionary phases, as noted by \citet{chaname2022mass}, reflects the fact that today's field red giants and the progenitors of today's field RC giants are not the same population---the latter being younger and extending to higher stellar masses---motivating an independent assessment of the binary fraction in both populations.

If binary interaction plays a differential role depending on the evolutionary stage, then a single combined sample could wash out or misattribute any real signal between the two populations. Testing the binary hypothesis independently in the RC and in the RGB is therefore essential to properly disentangle the origin of Li enrichment.

Observationally, the binary-enrichment hypothesis has been tested through radial-velocity (RV) monitoring campaigns of varying scales, precision, and time baselines, with results that remain inconclusive. Small early samples, such as \citet{1996A&A...309..465D} (12 stars) and \citet{2011ApJ...743..107R} (eight stars observed at only two epochs), did not find an enhanced binary frequency among Li-rich giants. \citet{2020A&A...639A...7J} monitored 11 Li-rich giants and 13 Li-normal giants over nine years, finding a binary frequency compatible with a normal reference sample of field K giants and no correlation with circumstellar dust, although most Li-enriched giants showed a possible link with fast rotation. \citet{2020MNRAS.498.2295G} monitored 18 stars (17 of them Li-rich) and found RV variation consistent with binarity for about a third of the sample, although with too few epochs to draw firm conclusions. Using indirect binarity proxies from \textit{Gaia} astrometry and RV uncertainties instead of dedicated monitoring, \citet{2024ApJ...964...42S} also found no significant excess of binarity indicators among Li-rich giants relative to a Li-normal sample ($\sim$15\% binary candidate rate in both populations).

\citet{2024A&A...690A.367C} previously compared Li-rich and Li-normal giants separately in each evolutionary phase. Combining RV epochs from \textit{Gaia}, the RAdial Velocity Experiment (RAVE), and GALAH, they found a slightly enhanced binary fraction among RGB Li-rich giants relative to Li-normal stars at the same phase, but no such enhancement for RC giants. Their multi-survey time baseline is sensitive to longer orbital periods, but at the cost of lower RV precision. More recently, \citet{sayeed26} carried out a dedicated RV follow-up of 58 Li-rich giants with Echelle SPectrograph for Rocky Exoplanet and Stable Spectroscopic Observations (ESPRESSO), a similar strategy to ours, and reported a companion fraction consistent with that expected for a background population of giants observed with the same cadence and precision. All of their candidates have $\log g = 2\text{-}3$ dex and A(Li)$\lesssim 2.7$ dex, although their sample was not explicitly designed to test the dependence on evolutionary stage.

These studies leave open whether the binary fraction of Li-rich giants truly depends on the evolutionary phase. In this work we adopt a complementary strategy: we combine high-precision, dedicated ESPRESSO monitoring with separate samples of Li-rich giants in the RC and the RGB, allowing us to test the binary-enrichment scenario independently in each phase.

This paper is organized as follows. We present the selection of our total sample of Li-rich giants, as well as the subsample selected for our program, in Sect. \ref{sec:data}. In Sect. \ref{sec:methods} we present the methods used to establish binary candidates, and describe the Monte Carlo simulations carried out to determine whether a candidate is consistent with a stellar binary. Our results are presented in Sect. \ref{sec:results}, and discussed in Sect. \ref{sec:discussion}. Finally, our conclusions are presented in Sect. \ref{sec:conclusion}.

\section{Data} \label{sec:data}
\subsection{Data selection} \label{subsec:2.1}
\begin{figure}[t!]
   \centering
 \includegraphics[width=\hsize]{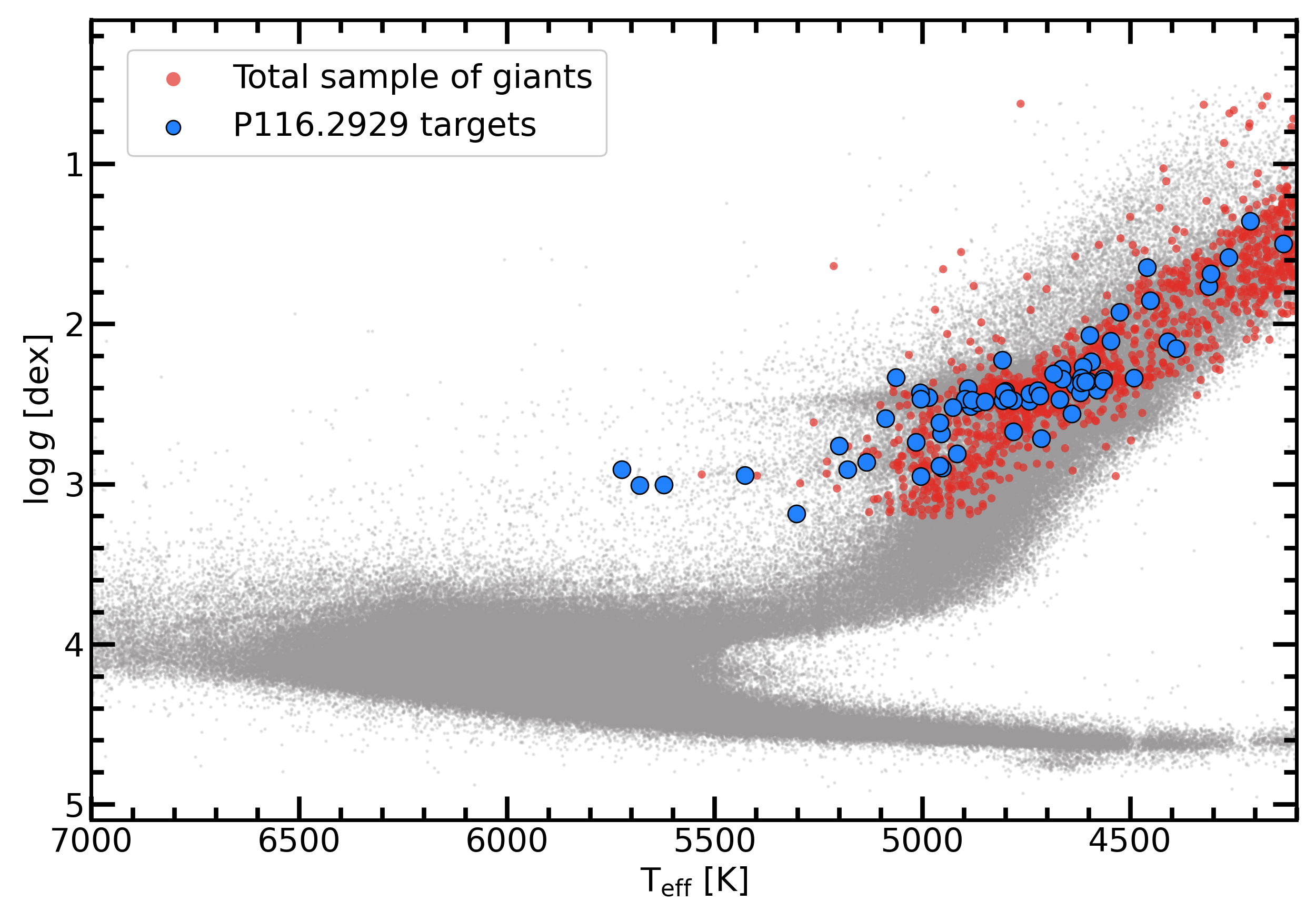}
 \caption{Kiel diagram showing our initial selected giants (red) and our 70 ESPRESSO targets analyzed in this work (blue). In gray we show the dataset from the GALAH DR3 survey.}
 \label{fig:Kiel}
\end{figure}

The total sample of 70 stars (40 RC and 30 RGB) observed and analyzed in this work was constructed first by cross-matching \textit{Gaia} Data Release (DR)3 \citep{2023A&A...674A...1G} with GALAH Data Release (DR)3 \citep{2017MNRAS.465.3203M, 2018MNRAS.478.4513B} to obtain the Li measurements. The selection of RC and RGB stars was then made following the same recipe used by \citet[][]{2021MNRAS.505.5340M}. This same procedure was applied in detail by \citet{2024A&A...690A.367C}, from whose resulting sample of Li-rich giants our targets were ultimately drawn; we refer the reader to their Sect. 3 for a full description of the quality flags and cross-match steps, and summarize only the key points here. GALAH \citep[][]{2015MNRAS.449.2604D} is a high-resolution (R$\sim$28,000) spectroscopic survey that observes stars with HERMES on the Anglo-Australian Telescope, providing stellar parameters and abundances for 30 elements for 588,571 stars. After applying data-quality cuts (based on the flags reported in GALAH), we restricted the sample to giant stars with effective temperatures 3000 $\leq\mathrm{T_{eff}}\leq$ 5730 K and surface gravities $\mathrm{Log\; g}\leq$ 3.2 dex, which enclose the expected range for red giants in the first-ascending RGB and the He-core burning phase. In Figure~\ref{fig:Kiel} we show this sample of giants and our 70 ESPRESSO targets on a Kiel diagram. We then defined a star as Li-rich if A(Li) $\equiv$ [Li/Fe] + [Fe/H] + 1.05 $>$ 1.5 dex, following the standard convention in the literature. Finally, we used the Bayesian stellar parameters estimator \texttt{(BSTEP)} isochrone-based evolutionary-state probability \citep{2018MNRAS.473.2004S} and the Wide-field Infrared Survey Explorer (WISE) \textit{$W_2$} absolute magnitude to distinguish the stellar evolutionary phase of the giants (RC or RGB) in GALAH DR3. This classification was validated against independent asteroseismic classifications from the \(\textit{Kepler}\) extended mission \citep[K2, ][]{2018MNRAS.476.3233H} for the subset of stars with available seismic data.

This selection procedure yielded a sample of 180 Li-rich giant stars that are included in the larger catalog of 1262 Li-rich giants identified by \citet[][]{2021MNRAS.505.5340M}. From this intersecting parent sample, we selected a subsample of 70 targets for our program, primarily constrained by brightness (9~$\leq\mathrm{V}\leq$~13.4) and visibility during the P116.2929 semester (hereafter P116). This sample size was chosen to provide sufficient statistical constraints to precisely determine and compare binary fractions within both the RC and RGB populations. If binary interactions are indeed the primary mechanism driving lithium enrichment, the binary fraction in our sample should be significantly elevated compared to the typical $\sim 30\%$ baseline observed in field giants \citep{2010ApJS..190....1R}.

Because the \texttt{BSTEP}-based classification does not explicitly distinguish the asymptotic giant branch (AGB), which can overlap with the RGB and RC in the Kiel diagram, our sample may contain a small number of misclassified AGB stars. \citet{2024A&A...690A.367C} estimated, from the relative evolutionary timescales of these phases, that AGB stars represent $\sim$1\% of their sample, and adopting the $\sim$5\% limit for AGB contamination among Li-rich giants reported by \citet{2019MNRAS.484.2000D}, up to $\sim$5\% of their 180 Li-rich giants could belong to the AGB. Scaled to our subsample of 70 targets, this translates into an upper limit of about three to four potentially misclassified AGB stars. We did not apply any additional method to remove such candidates, but given this small expected contamination, we do not expect it to affect our main conclusions regarding binarity and Li enrichment.

To ensure that our program is sensitive to the relevant binary configurations, we consider the theoretical period limits of the enrichment mechanism. For example, \citet{2019ApJ...880..125C} modeled a representative system consisting of a $1.5\,M_\odot$ giant and a $1\,M_\odot$ dwarf, and calculated the range of orbital periods capable of producing observable Li enrichment in the giant, deriving a minimum orbital period of 279 days by the need to avoid mass transfer through Roche lobe overflow, and an upper period limit of $\sim$3000 days for an effective mechanism of Li  enrichment due to tidal interactions. Consequently, any robust statistical test of the binary fraction must maintain sensitivity to periods up to this threshold.

Furthermore, we accounted for the presence of RV jitter in giant stars during our target selection. Specifically, \citet{Hekker2006} demonstrated that for giants with $\mathrm{B - V} \leq 1.2$, the intrinsic RV jitter remains below $\sim 50\ \mathrm{m\ s^{-1}}$. Therefore, we restricted our final sample to stars that meet this $\mathrm{B - V} \leq 1.2$ color criterion to ensure that stellar jitter does not mask or mimic the expected companion-induced Keplerian signals.

\subsection{ESPRESSO data}
All of our spectroscopic data were obtained with ESPRESSO \citep[][]{2021A&A...645A..96P}. ESPRESSO is a fiber-fed, high-resolution echelle spectrograph operating with $R \sim 70,000$---$140,000$, mounted on the 8.2 m Very Large Telescope (VLT) at the European Southern Observatory (ESO) in Cerro Paranal, Chile. Although designed to detect Earth-mass planets within the habitable zone of Sun-like stars by reaching an RV precision of down to $10\ \mathrm{cm\ s^{-1}}$ over baseline timescales of several years, its stability makes it ideal for characterizing stellar binaries. The observations were carried out as part of the P116 program. At the time of writing, the program had observed $\sim 82\%$ of the RC and $\sim 90\%$ of the RGB stars in our sample. We requested six RV epochs per target spanning the 7-month duration of the program; this observational cadence ensures high sensitivity across a broad range of periods and companion masses.

\begin{table*}
\centering
\caption{Properties and radial velocity data for a subset of the 57 targets.}
\label{tab:rv_epochs}
\footnotesize
\begin{tabular}{ccccccc}
\hline\hline
Main ID & Class & $A$(Li) & $\log g$ & BJD $-$ 2\,460\,000 & RV [km\,s$^{-1}$] & $\sigma_{\mathrm{RV}}$ [m\,s$^{-1}$] \\
\hline
TYC 9270-461-1    & RGB & $2.56$ & $2.07$ & [1083.79, 1110.76, 1130.77, ...] & [$-22.14$, $-22.14$, $-22.16$, ...] & [12, 8, 10, ...]  \\
TYC 7832-1535-1   & RGB & $3.18$ & $2.27$ & [1122.90, 1148.79, 1168.80, ...] & [$-6.51$, $-6.59$, $-6.55$, ...]    & [8, 7, 8, ...]    \\
TYC 5421-547-1    & RGB & $2.42$ & $2.11$ & [1010.79, 1033.83, 1055.78, ...] & [$39.75$, $39.80$, $39.83$, ...]    & [13, 11, 13, ...] \\
UCAC2\,12425336   & RC  & $2.15$ & $2.91$ & [1103.88, 1124.59, 1149.55, ...] & [$-14.31$, $-14.43$, $-14.39$, ...] & [15, 10, 13, ...] \\
UCAC4\,186-010575 & RC  & $1.66$ & $2.48$ & [1007.63, 1030.74, 1050.85, ...] & [$54.95$, $55.12$, $55.29$, ...]    & [7, 6, 7, ...]    \\
UCAC2\,23785777   & RC  & $2.82$ & $2.40$ & [981.64, 1006.64, 1027.56, ...]  & [$83.51$, $83.47$, $83.53$, ...]    & [9, 10, 9, ...]   \\
...   & ...  & ... & ... & ...  & ...    & ...   \\
\hline
\end{tabular}
\tablefoot{The full table is only available in electronic form at the Centre de Données astronomiques de Strasbourg (CDS).}
\end{table*}

To estimate the required exposure times, we utilized the ESPRESSO Exposure Time Calculator. We targeted a baseline S/N of 15 per pixel at $550\ \mathrm{nm}$, which yields an expected RV precision of $\leq 5\ \mathrm{m\ s^{-1}}$. Individual exposure times were customized based on the target magnitude, instantaneous seeing, airmass, and the specific instrumental configuration, ranging from $70\ \mathrm{s}$ for the brightest targets ($V = 11$) to $540\ \mathrm{s}$ for the faintest ($V = 13.4$).

The RV measurements were derived by processing each exposure with the ESPRESSO Data Reduction Software (DRS, version 3.3.19) executed via \texttt{EsoRex}. For each target, we provided the \textit{Gaia} DR3 catalog RV as a prior to center the cross-correlation window, along with the numerical stellar mask corresponding to the spectral type reported in GALAH DR3. The DRS cross-correlates the observed spectrum with this mask, and the final RV is determined by fitting a Gaussian profile to the resulting cross-correlation function. The final reduced spectra achieved an average S/N at $550\ \mathrm{nm}$ ranging between 5 and 35 per exposure ($\mu = 15$), with nominal RV uncertainties between $4$ and $78\ \mathrm{m\ s^{-1}}$.

Finally, to ensure the robustness of our binarity analysis, we filtered the data by excluding any target with fewer than four valid epochs, as well as individual observations assigned a ``C'' quality grade by the ESO pipeline. This quality cut resulted in a final science sample of 25 RGB stars and 32 RC, whose individual epochs and radial velocities are presented in Figures~\ref{fig:RGB_RVS} and \ref{fig:RC_RVS}, respectively. Table~\ref{tab:rv_epochs} illustrates a representative subset of the total sample, listing the main identifier, evolutionary stage, $A$(Li), $\log g$, and the first three epochs of the derived radial velocities and their associated uncertainties. The full table, including stellar parameters and all epochs for each target, is only available in electronic form at the CDS.

\section{Methods} \label{sec:methods}
To test binary interaction theory, we analyzed the final science sample of 57 targets through a two-step approach. First, we identified stars with statistically significant RV variability that is inconsistent with stellar jitter (Sect.~\ref{subsec:3.1}). And second, for the confirmed binary candidates, we constrained the companion masses and established the overall binary fraction of the total sample (Sect.~\ref{subsec:3.2}).

\subsection{Stellar jitter}\label{subsec:3.1}
The RV measurements of stars (especially evolved ones) are affected not only by the presence of orbiting companions but also by intrinsic stellar variability, commonly referred to as stellar jitter \citep{1997ApJ...485..319S}. This jitter arises from a combination of physical mechanisms acting on different timescales \citep{hatzes2019doppler}, including solar-like oscillations (minutes to days), granulation (hours to days), and rotational surface activity such as spots (days to months), among others.

\begin{figure*}
  \sidecaption
  \includegraphics[width=12cm]{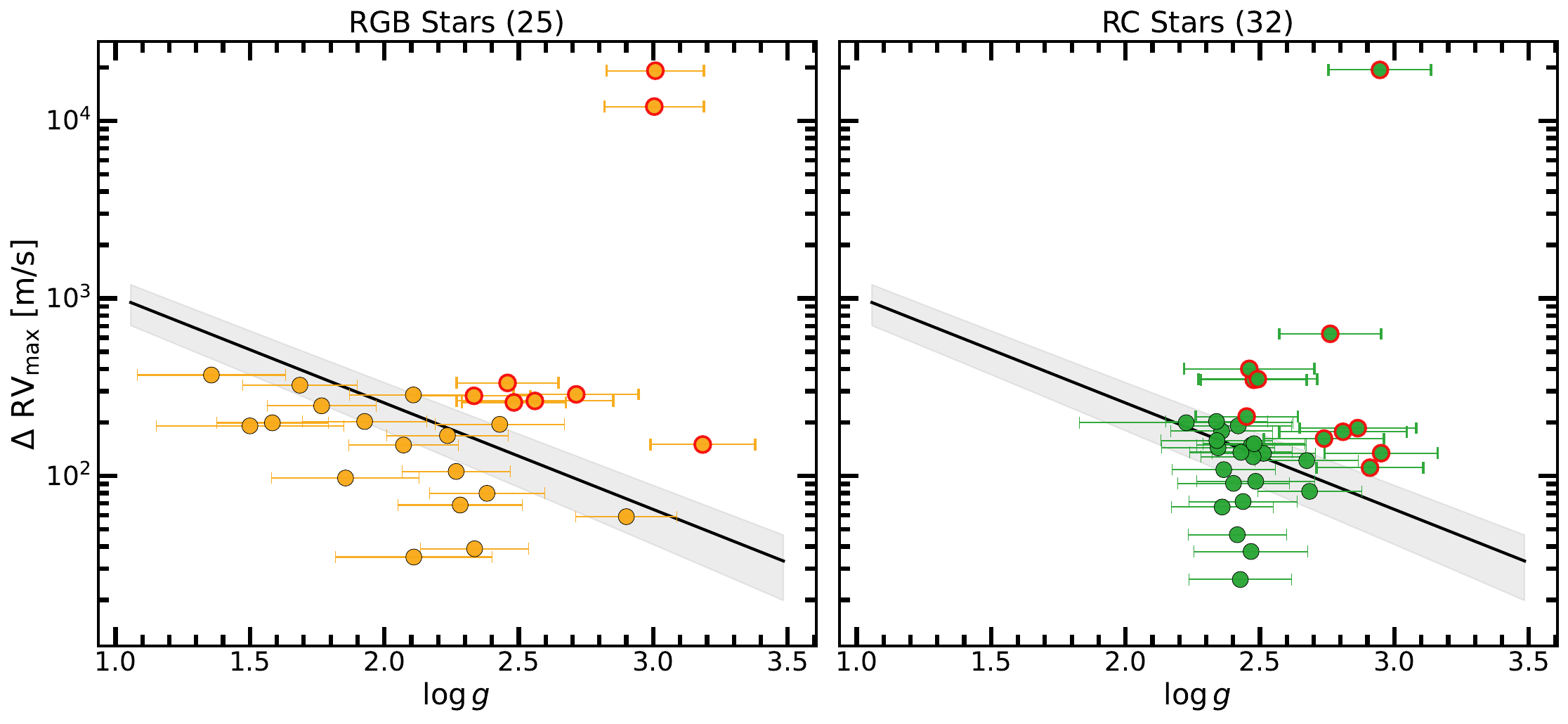}
  \caption{Observed peak-to-peak RV amplitude as a function of spectroscopic surface gravity for RGB (left) and RC (right) stars in our sample. The solid black line and shaded band show the median and $1\sigma$ uncertainty of the intrinsic jitter amplitude expected from the \citet{2008A&A...480..215H} relation. Horizontal error bars show the $1\sigma$ uncertainty in spectroscopic $\log g$ of our targets. Stars outlined in red are classified as companion candidates according to our criterion (Sect.~\ref{subsec:3.1}).}
  \label{fig:hekker}
\end{figure*}

In red giants, oscillation-driven jitter is expected to dominate and scale with the star evolutionary state \citep{1995A&A...293...87K}, reaching amplitudes of several to tens of $\mathrm{m\ s^{-1}}$ as stars evolve up the giant branch. This is why in addition to our first color criterion found by \citet{Hekker2006} (previously mentioned in Sect. \ref{subsec:2.1}), we adopted the empirical relation between nonperiodic RV variation ($\mathrm{v_r}$) and surface gravity, originally found by \citet{2008A&A...480..215H} and later parametrized by \citet{2016AJ....151...85T}. Following the log-log formulation of \citet{2018A&A...620A.139L}
\begin{equation}
    \mathrm{log\;(v_r[m/s]) =}\; a\cdot \mathrm{log}\;g + b \label{ec1}
\end{equation}

with coefficients $a=-0.60$$\,\pm\,0.04\,$dex$^{-1}$ and $b=3.31$$\,\pm\,0.10$ \citep[also adopted by \citealp{sayeed26}]{2008A&A...480..215H}. Therefore, for each of our 57 targets we defined the observed RV variation as the full range between its maximum and minimum measured velocity, $\mathrm{\Delta RV_{max} = RV_{max} - RV_{min} }$, with the associated uncertainty propagated from the individual errors of these two epochs.

Since this relation (Equation \ref{ec1}) predicts a semi-amplitude while $\mathrm{\Delta RV_{max}}$ is a peak-to-peak range (in the most extreme scenario), we scaled the predicted value by a factor of two before comparison. The uncertainties in $a$, $b$, and $\log g$ are propagated to the prediction assuming independent errors and first-order (linear) error propagation, which gives
\begin{equation}
    \sigma_{v_r} \simeq v_r \,\ln(10)\,
    \sqrt{(\log g)^{2}\,\sigma_{a}^{2} + \sigma_{b}^{2} + a^{2}\,\sigma_{\log g}^{2}} .
    \label{eq:sigma_v}
\end{equation}
The resulting $v_r \pm \sigma_{v_r}$, evaluated on the full $\log g$ grid and scaled by the same factor of two, defines the shaded band in Figure~\ref{fig:hekker}. As a check, we compared this analytic estimate against a Monte Carlo realization sampling a, b, and $\log g$ from their respective Gaussian uncertainties. The two approaches yield statistically consistent bands, with no changes in classification (companion candidate vs. non-companion candidate) among the 57 stars in our sample.

Finally, a star is flagged as a companion candidate if its observed $\mathrm{\Delta RV_{max}}$ exceeds the 1$\sigma$ upper bound of this predicted jitter distribution. We selected this threshold to be more permissive than the 3.5$\sigma$ criterion adopted in the literature \citep{2008A&A...480..215H, sayeed26}, as candidates are subsequently vetted through orbital characterization (Sect. \ref{subsec:3.2}), which independently discards solutions inconsistent with a stellar, substellar, or a planetary companion. Under this criterion, 19/57 stars ($33 \pm 6$\%) are classified as binary candidates or stars with a companion, with comparable fractions among RC (11/32, $34 \pm 8$\%) and RGB (8/25, $32 \pm 9$\%) stars. The uncertainties quoted are binomial errors. Figure~\ref{fig:hekker} shows the observed $\mathrm{\Delta RV_{max}}$ as a function of $\log g$ for both evolutionary stages with the expected jitter relation and its uncertainty band, where stars outlined in red correspond to the companion candidates identified by our criterion.

We note that additional RV epochs from large-scale surveys such as \textit{Gaia} or GALAH could, in principle, be used to extend our time baseline and further test the jitter and/or binarity classification. However, combining RV epochs from ESPRESSO with those from \textit{Gaia} or GALAH requires accounting for zero-point offsets between instruments, which are nonnegligible relative to the RV variations relevant to our analysis and could therefore introduce spurious signals or reduce the precision of the jitter and/or binarity assessment. Therefore, we chose to rely exclusively on the ESPRESSO data and defer this type of multi-survey combination to future work, once \textit{Gaia} DR4 becomes available and provides a more robust framework for combining RV data across instruments.

\subsection{Companion masses}\label{subsec:3.2}
To determine whether an unresolved stellar companion is present in the 19 previously filtered targets, we used the Monte Carlo simulation developed by Vilaza-Dallago et al. (in prep.). This method combines the observed RV amplitude with the stellar mass of the primary to derive the possible periods and masses that a companion would need to explain the observations. This gives an indication of the type of object (star, brown dwarf, or planet) that could produce the observed variability. We briefly describe the method here, and a full description of the simulation and its validation will be presented in that work.

For each target we constructed a logarithmic grid in companion mass ($M_2$, spanning $\sim 1M_{jup}$ up to the mass of the primary) and orbital period ($P_{orb}$, spanning 1 up to 3000 days), with a grid resolution of 40$\times$40 points. Then, for each ($M_2, P_{orb}$) pair, synthetic RV curves were generated using the \texttt{RadVel} package \citep{2018PASP..130d4504F}. For each curve, we sampled the orbital inclination ($\cos{i}$, uniformly in $[-1,1]$) and the argument of periastron ($\omega$, uniformly in [0,2$\pi$]), following the same scheme as Vilaza-Dallago et al. (in prep.). In contrast to that work, which adopts a uniform eccentricity distribution in $[0,0.9]$, we used the eccentricity prior of \citet{2013MNRAS.434L..51K}, modeled as a Beta distribution
\begin{equation}
    P_{\beta}(e; c,d) = \mathrm{Beta}(c,\;d) = \frac{\Gamma(c+d)}{\Gamma (c) \Gamma (d)}\;e^{c-1}\;[1-e]^{d-1} \label{ec2}
\end{equation}

with $c = 0.867,\; d=3.03$, which better reflects the observed eccentricity distribution of binary and/or multi-planet systems. For each ($M_2, P_{orb}$) pair we also generated 30 distinct observation-phase realizations with 30 orbital-parameter draws, yielding a total of 900 synthetic curves per pair.

Furthermore, for each ($M_2, P_{orb}$) grid point, a combination was considered consistent with the data if the observed amplitude fell within 1$\sigma$ of the simulated amplitude distribution, computed after adding an instrumental jitter of 50 m s$^{-1}$ in quadrature to the simulated amplitudes to ensure a consistent noise budget between the simulated and observed data. Finally, to verify that our results were not sensitive to a single random realization of Monte Carlo sampling, we repeated this entire procedure with 20 independent seeds per star and confirmed that the set of filtered ($M_2, P_{orb}$) pairs remained stable across seeds.

In addition to this last criterion, since the primary radius $R_1$ can be significantly large for some stars (specially for evolved RGB), this may result in an attached binary system inducing Roche lobe overflow. Therefore, for each target, we calculated, using the \citet{1983ApJ...268..368E} approximation, the effective Roche lobe radius of the primary as a function of the mass ratio $q=M_1/M_2$
\begin{equation}
    \frac{R_L}{r} = \frac{0.49\,q^{2/3}}{0.6\,q^{2/3}+\ln(1+q^{1/3})},
    \label{eq:eggleton}
\end{equation}
where $r$ is the orbital semi-major axis. Setting $R_L = R_1$ and combining with Kepler's third law, we solved for the minimum orbital period $P_{orb, min}$ below which the primary would fill its Roche lobe, as a function of $M_2$. Given that we require the system to remain detached in order for Li enrichment to proceed via tidal locking or tidal interaction, we exclude the grid points with $P_{orb}<P_{orb,min}$.

This process yields a grid of ($M_2, P_{orb}$) pairs that satisfy both the amplitude-consistency and Roche lobe criteria, and thus reproduce the observed RV variability under physically plausible configurations. From this filtered grid, we estimate the probability that a target hosts a stellar companion ($P_{SB}$) as the fraction of grid points with $M_2$ above the minimum mass required for a star to sustain core hydrogen fusion, set at 80 $M_{jup}$ \citep{2015ApJ...810L..25H}. We also define $P_{SS}$ as the fraction of grid points with $M_2$ above the deuterium-burning limit, set at 13 $M_{jup}$ that is commonly adopted as the boundary between planets and brown dwarfs \citep{2001RvMP...73..719B}. In contrast to Vilaza-Dallago et al. (in prep.), who applied a cut and classified a target as a stellar binary if $P_{SB} > 0.67$, we report the full $P_{SB}$ value as well as $P_{SS}$ value for each target without discarding possible solutions.

\section{Results} \label{sec:results}
\begin{figure*}[ht!]
  \centering
  \includegraphics[width=1.0\textwidth]{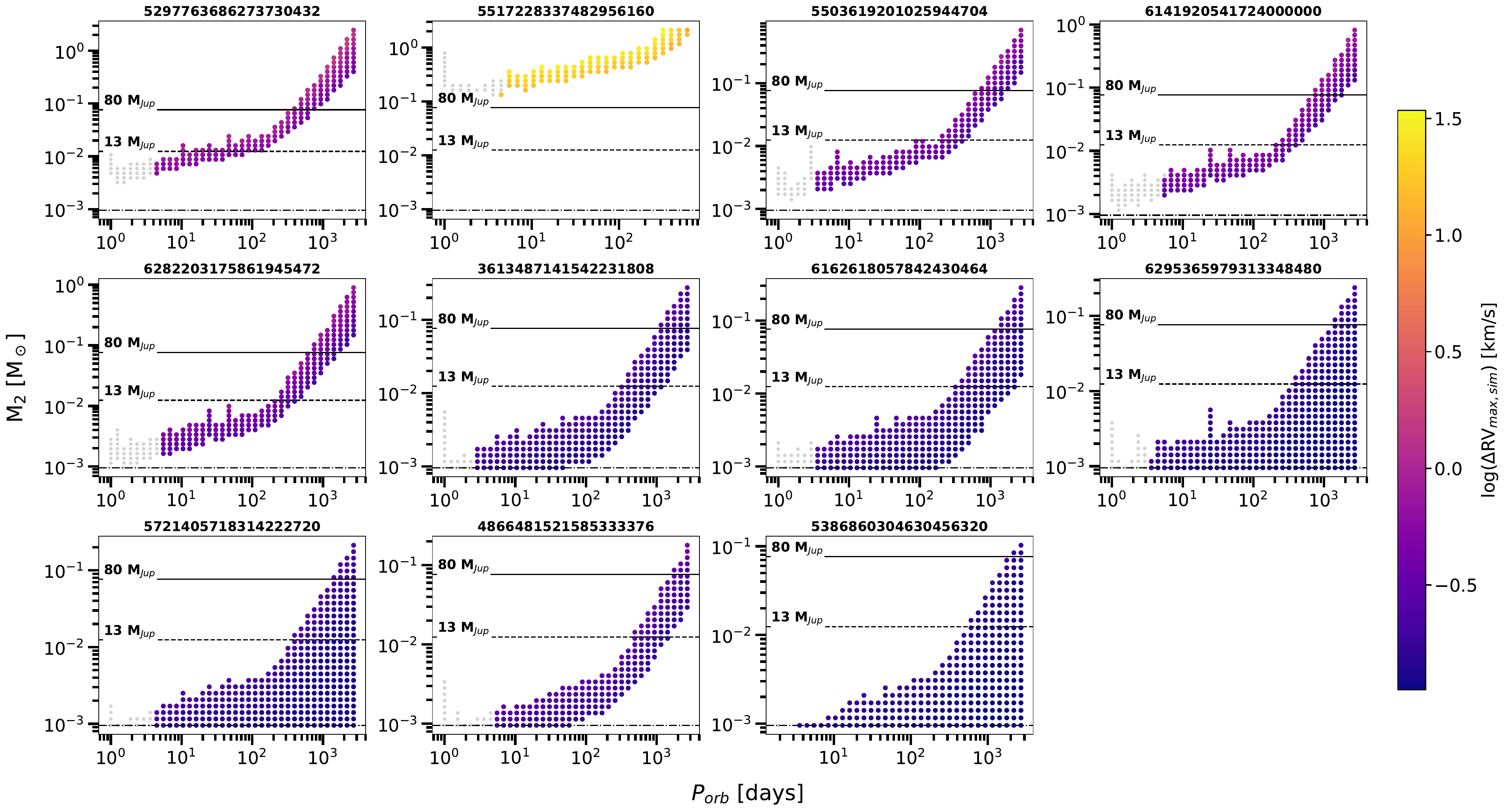}
  \caption{Simulation results for the 11 RC targets, identified by their \textit{Gaia} DR2 source ID. The solid line shows the limit considered between a star and a brown dwarf (80 $M_{jup}$), the dashed line shows the limit between a brown dwarf and an exoplanet (13 $M_{jup}$), and gray points indicate grid points excluded by the Roche lobe criterion. Each dot is colored by the simulated RV amplitude of its distribution.}
  \label{fig:RC_model}
\end{figure*}

\begin{figure*}[ht!]
  \centering
  \includegraphics[width=1.0\textwidth]{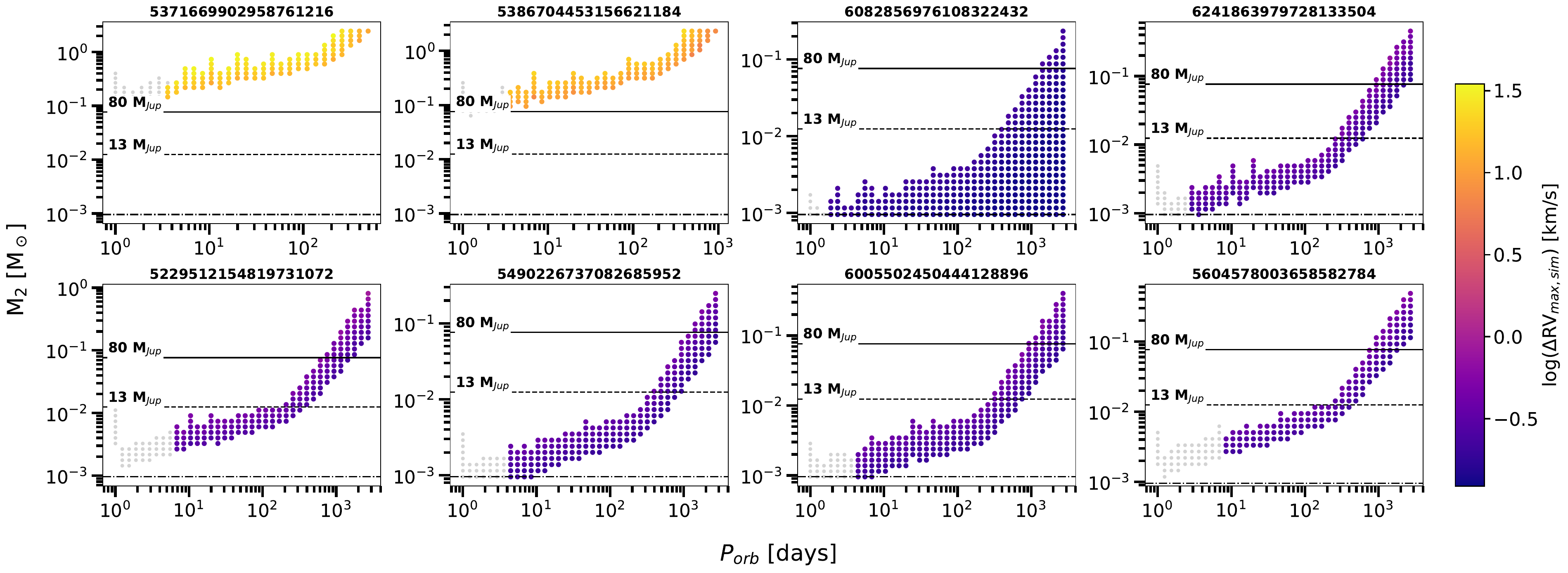}
  \caption{Simulation results for the eight RGB targets, identified by their \textit{Gaia} DR2 source ID. The solid line shows the limit considered between a star and a brown dwarf (80 $M_{jup}$), the dashed line shows the limit between a brown dwarf and an exoplanet (13 $M_{jup}$), and gray points indicate grid points excluded by the Roche lobe criterion. Each dot is colored by the simulated RV amplitude of its distribution.}
  \label{fig:RGB_model}
\end{figure*}

Following the procedure described in Sect. \ref{subsec:3.2}, we ran the simulation for the 11 RC and eight RGB targets that survived the filtering criteria of Sect. \ref{subsec:3.1}. The results for each sample are shown in Figures~\ref{fig:RC_model} and~\ref{fig:RGB_model}, for the RC and RGB samples, respectively. Each panel displays the ($M_2, P_{orb}$) grid points that the model can reproduce given the observed RV variation, with the color bar indicating the simulated RV amplitude on a logarithmic scale. The two mass thresholds discussed above, 80 $M_{jup}$ (stellar companion) and 13 $M_{jup}$ (substellar companion), are marked in each panel by a solid and a dashed line, respectively. The grid points excluded by the Roche lobe criterion (Sect. \ref{subsec:3.2}) are shown in gray and are concentrated toward the left side of each panel, at short periods.

As expected for this type of system, both figures show a clear trend in which the minimum companion mass required to reproduce the observed RV variability increases with orbital period. The most notable exceptions are three targets (two RGB and one RC) for which the entire ($M_2, P_{orb}$) grid falls within the stellar-companion regime (yellow points in Figures~\ref{fig:RC_model} and~\ref{fig:RGB_model}).

To quantify these plots across the 19 targets, we computed the fraction of grid pairs above each mass threshold, $P_{SB}$ (stellar companion) and $P_{SS}$ (substellar companion) for every target. Table \ref{tab:Results_table} summarizes for each group of stars, the number and percentage of targets whose values $P_{SB}$ and $P_{SS}$ exceed 0.67, 0.5, and 0.25.

\begin{table}
\caption{Number and percentage of targets exceeding the companion probability thresholds.}
\label{tab:Results_table}
\centering
\begin{tabular}{lccc}
\hline\hline
 & $P>0.67$ & $P>0.5$ & $P>0.25$ \\
\hline
\multicolumn{4}{l}{RGB ($N = 8$)} \\
$P_{SB}$ & 2 (25.0\%) & 2 (25.0\%) & 3 (37.5\%) \\
$P_{SS}$ & 2 (25.0\%) & 4 (50.0\%) & 7 (87.5\%) \\
\hline
\multicolumn{4}{l}{RC ($N = 11$)} \\
$P_{SB}$ & 1 (9.1\%) & 1 (9.1\%) & 4 (36.4\%) \\
$P_{SS}$ & 2 (18.2\%) & 4 (36.4\%) & 8 (72.7\%) \\
\hline
\end{tabular}
\tablefoot{Targets in each sample (RC and RGB) whose stellar companion probability ($P_{SB}$) and substellar companion probability ($P_{SS}$) exceed the indicated thresholds (0.67, 0.5, and 0.25).}
\end{table}

\section{Discussion} \label{sec:discussion}
In our first criterion for RV jitter, we find that $34 \pm 8$\% of RC targets and $32 \pm 9$\% of RGB targets are consistent with the presence of a close companion. This first result is compatible with the expected binary fraction for a random sample ($\sim$30\%), and we find no significant differences between the RC and RGB populations.

These results can be directly compared with the analysis made by \citet{2024A&A...690A.367C}, who found that the binary fraction among RGB Li-rich giants was slightly enhanced relative to Li-normal stars at the same evolutionary phase, while no such enhancement was found for RC giants. In contrast, our results do not show a significant excess in either evolutionary phase. This difference likely reflects the complementary strengths and limitations of each approach, rather than a discrepancy in the underlying data. While our ESPRESSO monitoring provides significantly higher RV precision, it is also limited to an $\sim$7 month baseline, making our method mostly insensitive to long-period systems. \citet{2024A&A...690A.367C}, on the other hand, benefit from RV epochs spanning a much longer temporal baseline (spanning several years across \textit{Gaia}, RAVE, and GALAH), which in principle allows them to probe longer periods systems, at the cost of lower precision. 

A similarly normal binary fraction has been reported from long-baseline RV monitoring. \citet{2020A&A...639A...7J} carried out a nine-year RV follow-up of 11 Li-rich K giants, together with a control sample of 13 Li-normal giants, and found a binary fraction compatible with that of a reference sample of field K giants, with no correlation between Li enrichment and circumstellar dust excess. Despite the very different observing strategies---a nine-year, sparsely sampled baseline versus our seven-month, densely sampled, high-precision campaign---both studies converge on the same overall conclusion: Li-rich giants do not show an excess binary fraction relative to representative field populations.

Our results are also consistent with the survey-based analysis of \citet{2024ApJ...964...42S}, which found no significant differences in these binarity fraction proxies between the two populations. Building on that same parent sample, \citet{sayeed26} carried out a dedicated RV follow-up with ESPRESSO of 58 Li-rich giants over a $\sim$6-month baseline, closely matching our own observational strategy. They report an overall companion fraction of 17\% (10/58), consistent with the $\sim$16\% they estimate for normal giants observed with the same cadence and precision, in line with our finding of no enhanced companion fraction. All of their candidates have $\log g = 2\text{-}3$ dex and A(Li) $\lesssim2.7$ dex; however, the 26\% companion rate in that $\log g$ range is not statistically significant ($p\approx$ 12\%), and they estimate that core He-burning stars make up at most $\sim$24\% of it, so their data do not by themselves indicate an RC-specific enhancement. Note that the companion fractions are not directly comparable, since our jitter criterion (1$\sigma$ upper bound) is more permissive than theirs (5th-percentile semi-amplitude above 3.5$\sigma$). Our study, which purposely targets independent RC and RGB samples, finds statistically indistinguishable fractions consistent with the presence of a close companion between our RC ($34 \pm 8$\%) and RGB ($32 \pm 9$\%) targets. Together with the slight RGB excess reported by \citet{2024A&A...690A.367C}, the current observational picture regarding a possible evolutionary dependence of the binary fraction among Li-rich giants remains unsettled, confirming the need for larger, dedicated, and evolutionary-stage-resolved RV campaigns like ours.

Based on Figures~\ref{fig:RC_model} and~\ref{fig:RGB_model}, we confirm that three of our 19 binary candidates are stellar systems with companions of $M > 80M_{jup}$. Looking also at the observed peak-to-peak RV amplitude as a function of $\log g$ (Figure~\ref{fig:hekker}), we see that these three stars are located in the top right part of the plots, each having a $\Delta$RV$_{max}$ value exceeding 10 km s$^{-1}$. These results agree with the threshold established by \citet{2019ApJ...875...61M} for the detection of stellar binary systems, which is $\Delta$RV$_{max}$ = 1 km s$^{-1}$, confirming that these systems require a binary companion to explain their variations. We note that a confirmed companion fraction of 3/57 ($\sim$5\%) at the stellar-mass regime corresponds to a lack of the binary fraction expected for a random sample of field giants and therefore this result does not by itself support a scenario in which binarity is a necessary condition for Li enrichment.

Looking in more detail at Table \ref{tab:Results_table}, there is agreement between the fraction probabilities in both samples. Specifically, the observed RV variation can be explained by a stellar binary system with $P_{SB}>0.25$ ($\sim$37\%) and by a system with a substellar companion with $P_{SS}>0.25$ ($\sim$80\%), with no significant difference found between the two stellar populations.

Our result is also directly relevant to the theoretical prediction of \citet{2019ApJ...880..125C}, who argued that tidally induced spin-up during the core-helium flash should synchronize the majority of RC giants with a binary companion, thereby driving the extra mixing required for Li production. If binary interactions were the dominant mechanism responsible for producing Li-rich giants, we would expect a clear excess of binary candidates among our RC targets relative to the $\sim$30\% field binary fraction reported by \citet{2010ApJS..190....1R}. Instead, we find $P_{SB}$ values that are statistically indistinguishable between the RC and RGB samples, and from the field binary fraction. This is consistent with the lack of enhanced binarity reported by \citet{2024A&A...690A.367C} for RC Li-rich giants, and argues against binary interaction being the main pathway of Li enrichment for the general population of Li-rich giants.

There are also four targets across both samples for which we cannot find a significant probability of a stellar or substellar companion (three in the RC sample and one in the RGB sample). Not surprisingly, these are the targets in Figures~\ref{fig:RC_model} and~\ref{fig:RGB_model} that have most of their grid points in the lower zone; more specifically, they are the ones for which, even in the most extreme scenario ($P_{orb}$ = 3000 days), the RV variation can be explained by a companion mass of 1$M_{jup}$. These are probably targets whose RV variation is at the limit of what can be explained by RV jitter.

Although the lack of an excess binary fraction in either evolutionary phase does not rule out binary interactions for individual Li-rich giants, it indicates that binarity is not the primary driver of Li enrichment for the population as a whole, at least for the orbital periods probed here. Finally, we emphasize that while our initial sample comprises 57 targets, our candidate selection and orbital characterization are based on a $\sim$7-month observing baseline; these results should therefore be interpreted with caution regarding longer-period systems. Extending the RV baseline will be essential not only to confirm the 19 candidates currently flagged above the jitter threshold---and potentially uncover new ones as their observed $\Delta\mathrm{RV_{max}}$ grows---but also to probe wider orbits and cleanly disentangle binary-induced spin-up from internal enrichment mechanisms across the entire sample.

\section{Conclusions} \label{sec:conclusion}
We analyzed high-resolution spectra of 57 Li-rich giant stars obtained with the ESPRESSO spectrograph over a time span of $\sim$7 months, in order to detect possible RV variations in these stars and thereby explore whether the binary interaction theory can explain the origin of this type of star. More specifically, 32 RC and 25 RGB stars were studied to test whether the evolutionary stage of the star is directly correlated with the Li-enrichment mechanism. The first criterion applied in our analysis was to discard RV variations likely produced by stellar jitter, for which the stellar parameters obtained from the GALAH DR3 survey were essential. After this, only 11 ($34 \pm 8$\%) RC and eight ($32 \pm 9$\%) RGB stars were classified as consistent with the presence of a close companion. To establish whether these targets were spectroscopic binaries, we performed a Monte Carlo simulation using the observed RV variation together with the stellar properties of the targets to assess the presence of an unresolved companion. The simulations yielded a total of three out of 19 ($16 \pm 8$\% of the filtered sample) targets confirmed as stellar binaries. Together with the fact that the probabilities of a stellar companion ($P_{SB}$) and of a substellar companion ($P_{SS}$) are similar between the two samples (RC and RGB) for values above 0.25, we find no significant differences between the two evolutionary phases. This is not what would be expected if binary interaction were the main driver of the Li enrichment in one of the two phases, although our current baseline limits the strength of this statement. This lack of an excess binary fraction in either evolutionary phase suggests that, at least for the short orbital periods probed here, binary interaction is unlikely to be the dominant mechanism driving Li enrichment in these giants.

These results should be taken with caution given the preliminary nature of this analysis. Our ESPRESSO program is still ongoing, and 13 of the 70 proposed targets have not do yet had more than four observations. Most importantly, our current $\sim$7-month baseline restricts our sensitivity to relatively short orbital periods, leaving longer-period systems---the regime expected in tidal spin-up scenarios \citep[e.g.,][]{2019ApJ...880..125C}---unconstrained. Additional RV epochs extending the observational baseline would extend our sensitivity to these wider systems and are therefore essential before a definitive conclusion on the role of binarity in Li-rich giants can be reached. 

\section*{Data availability}
Table~\ref{tab:rv_epochs} is only available in electronic form at the CDS via anonymous ftp to \url{cdsarc.u-strasbg.fr} (130.79.128.5) or via \url{http://cdsweb.u-strasbg.fr/cgi-bin/qcat?J/A+A/}.

\begin{acknowledgements}
We thank the anonymous referee for their careful reading of the manuscript and for their constructive comments and suggestions, which have helped to improve the quality of this paper. J.C. acknowledges support from the Agencia Nacional de Investigación y Desarrollo (ANID) via Proyecto Fondecyt Regular 1231345, and by ANID BASAL project FB210003. The Geryon cluster at the Centro de Astro-Ingenieria UC was extensively used for the calculations performed in this paper. ANID BASAL project FB21000, BASAL CATA PFB-06, the Anillo ACT-86, FONDEQUIP AIC-57, and QUIMAL 130008 provided funding for several improvements to the Geryon cluster. 
\end{acknowledgements}

\bibliographystyle{aa}
\bibliography{sample631}

\begin{appendix}

\onecolumn
\section{Individual radial velocity time series}
\begin{figure*}[ht!]
  \centering
  \includegraphics[width=1.0\textwidth]{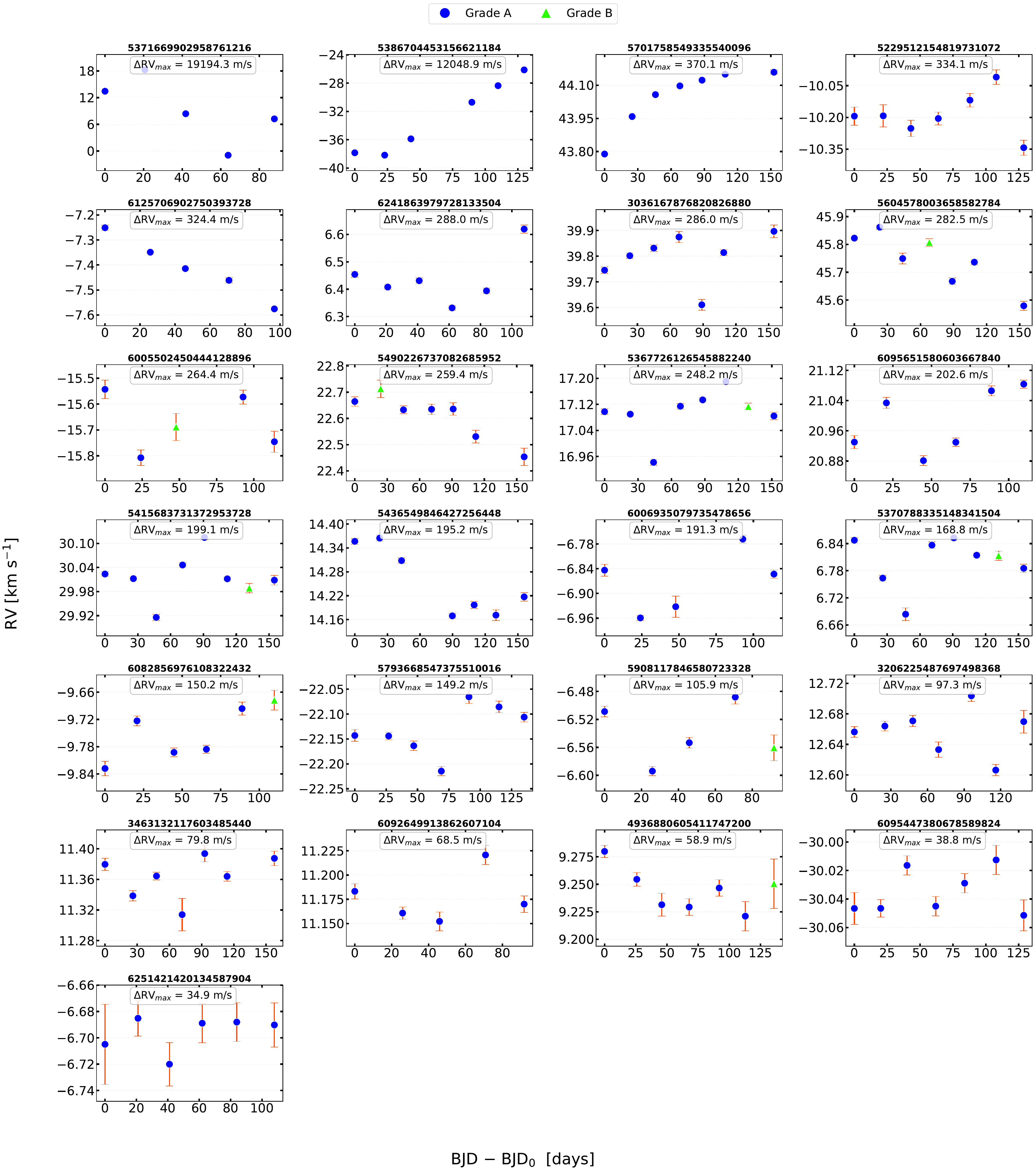}
  \caption{RV time series for the 25 RGB stars, ordered by decreasing $\Delta\mathrm{RV_{max}}$, from top-left to bottom-right, row by row. Each panel shows RV as a function of time ($\mathrm{BJD}-\mathrm{BJD_0}$, in days) with $1\sigma$ error bars; in several panels the error bars are smaller than the plotted symbols and are therefore not visible. Blue circles denote observations of quality grade A and lime triangles those of grade B; grade C observations were excluded from the analysis. The $\Delta\mathrm{RV_{max}}$ value in each panel is the maximum RV variation measured for that star.}
  \label{fig:RGB_RVS}
\end{figure*}

\begin{figure*}[ht!]
  \centering
  \includegraphics[width=1.0\textwidth]{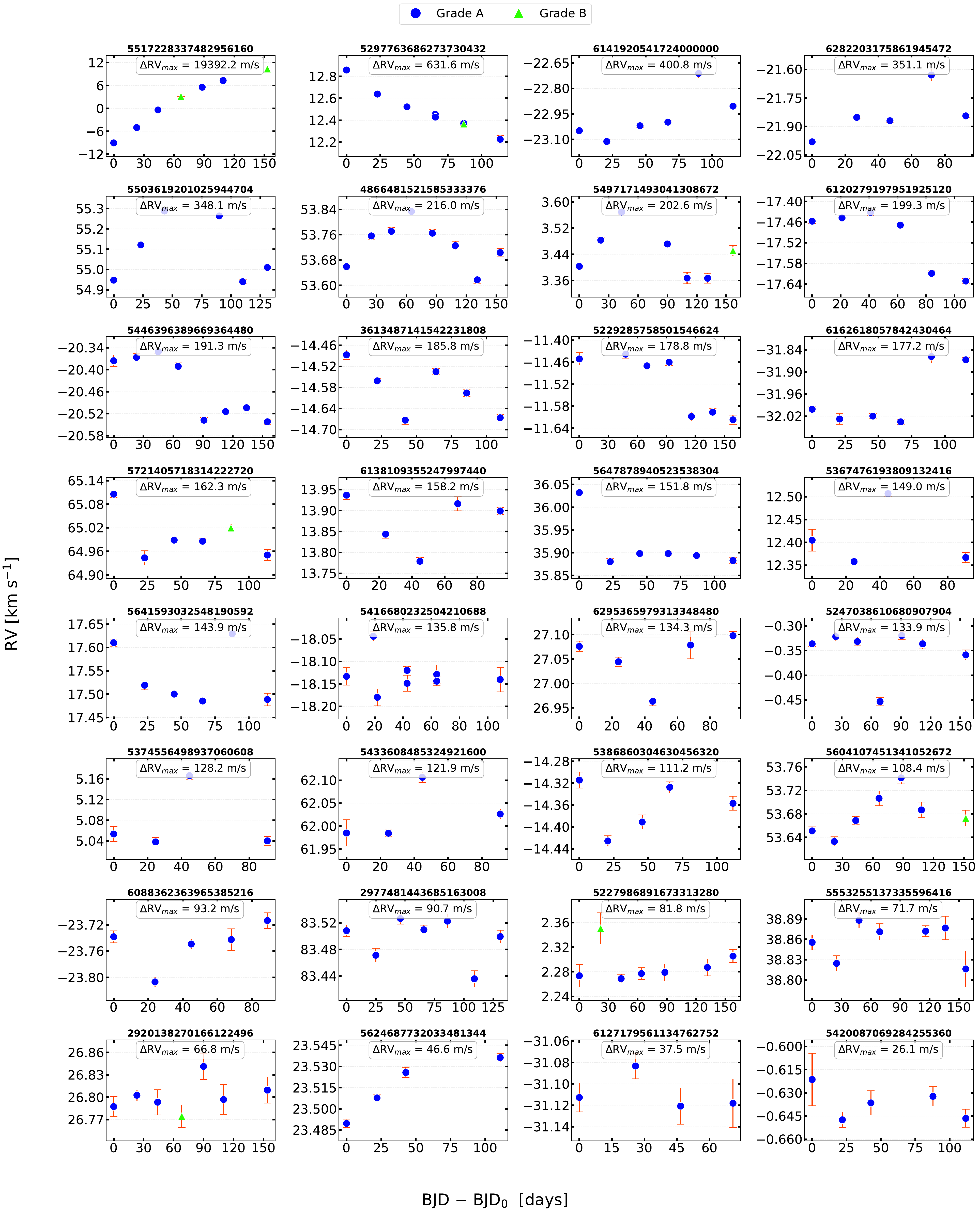}
  \caption{RV time series for the 32 RC stars, ordered by decreasing $\Delta\mathrm{RV_{max}}$, from top-left to bottom-right, row by row. The symbols, error bars, and quality grades are the same as in Figure~\ref{fig:RGB_RVS}.}
  \label{fig:RC_RVS}
\end{figure*}

Figures~\ref{fig:RGB_RVS} and \ref{fig:RC_RVS} show the RV time series of all RGB and RC stars, respectively.

\end{appendix}

\end{document}